\documentclass[aps,prl,preprint,showpacs,superscriptaddress,groupedaddress]{revtex4}
\usepackage{bm}
\usepackage{amsmath}
\usepackage{amsfonts}%
\usepackage{amssymb}%
\usepackage{graphicx}
\usepackage{color}
\usepackage{hyperref}
\usepackage{verbatim} 
\usepackage{caption}

\begin{document}
\newcommand{\gammaref}{\Gamma_{\mathrm{ref}}}
\newcommand{\omegaref}{\omega_{\mathrm{ref}}}
\newcommand{\homegaref}{\hat{\omega}_{\mathrm{ref}}}
\newcommand{\fref}{\tilde{f}_{\mathrm{ref}}}
\newcommand{\homega}{\hat{\omega}}
\newcommand{\meter}{\mathrm{m}}
\newcommand{\nm}{\mathrm{nm}}
\newcommand{\cpar}{\mathrm{C}}
\newcommand{\mdel}{m_{\delta}}
\newcommand{\hk}{\tilde{k}} 
\newcommand{\gam}{\dot\gamma}
\newcommand{\ndelta}{N_{\delta}}
\newcommand{\dgam}{\hat{\dot\gamma}}
\newcommand{\ar}{\mathrm{DC}}
\newcommand{\hx}{\hat{x}}
\newcommand{\hv}{\hat{v}}
\newcommand{\ha}{\hat{\ddot x}}
\newcommand{\ot}{\omega t}
\newcommand{\dom}{\Delta \omega}
\newcommand{\dof}{\Delta f}
\newcommand{\hs}{\hat{\sigma}}
\newcommand{\hsa}{\hat{\sigma}^{(a)}}
\newcommand{\hsm}{\hat{\sigma}_m}
\newcommand{\ii}{\mathrm{i}}
\newcommand{\impre}{\mathcal{Z}_{R}}  
\newcommand{\impim}{\mathcal{Z}_{I}}  
\newcommand{\imp}{\mathcal{Z}}  
\newcommand{\impa}{\mathcal{Z}^{(a)}}  
\newcommand{\impf}{\mathcal{Z}^{(0)}}  
\newcommand{\impfree}{\mathcal{Z}^{(\mathrm{L})}}  
\newcommand{\impaf}{\mathcal{Z}^{(a_\mathrm{fluid})}}  
\newcommand{\impdna}{\mathcal{Z}^{(\mathrm{DNA})}}  
\newcommand{\impldna}{\mathcal{Z}^{(\mathrm{LDNA})}}  
\newcommand{\vb}{v^{(0)}} 

\bibliographystyle{apsrev}

\title{Hydrodynamics of quartz crystal microbalance
  experiments with liposome-DNA complexes}
\author{Adolfo Vazquez-Quesada}
\affiliation{Departmento de Fisica de la Materia Condensada, Universidad Autonoma de Madrid,
and Institute for Condensed Matter Physics, IFIMAC.
Campus de Cantoblanco, Madrid 28049, Spain}
\author{Marc Melendez Schofield}
\affiliation{Departmento de Fisica de la Materia Condensada, Universidad Autonoma de Madrid,
and Institute for Condensed Matter Physics, IFIMAC.
Campus de Cantoblanco, Madrid 28049, Spain}
\author{Achilleas Tsortos}
\affiliation{Institute of Molecular Biology and Biotechnology, Foundation for Research and Technology-Hellas, Heraklion, Crete, 70013, Greece}
\author{Pablo Mateos-Gil}
\affiliation{Institute of Molecular Biology and Biotechnology, Foundation for Research and Technology-Hellas, Heraklion, Crete, 70013, Greece}
\author{Electra Gizeli}
\affiliation{Institute of Molecular Biology and Biotechnology, Foundation for Research and Technology-Hellas, Heraklion, Crete, 70013, Greece}\affiliation{Department of Biology, University of Crete, Heraklion 71110, Greece}
\author{Rafael Delgado Buscalioni}
\email[]{rafael.delgado@uam.es}
\affiliation{Departmento de Fisica de la Materia Condensada, Universidad Autonoma de Madrid,
and Institute for Condensed Matter Physics, IFIMAC.
Campus de Cantoblanco, Madrid 28049, Spain}

\begin{abstract}
The quartz crystal microbalance (QCM) is widely used to study surface adsorbed molecules, often of biological significance.  However, the relation between raw acoustic response (frequency shift $\Delta f$  and dissipation factor $\Delta D$) and mechanical  properties of the macromolecules still needs to be deciphered, particularly in the case of suspended discrete particles.   We study the QCM response of suspended liposomes tethered to the resonator wall by double stranded DNA, with the other end  attached to surface-adsorbed neutravidin through a biotin linker. Liposome radius and dsDNA contour length are  comparable to the wave penetration depth ($\delta\sim 100\ \mathrm{nm}$). Simulations, based on the immersed boundary  method and an elastic network model for the liposome-DNA complex, are in good agreement with experimental results for  POPC liposomes. We find that the added stress at the resonator surface, i.e. the impedance Z sensed by QCM, is dominated 
by the
flow-induced liposome surface-stress, which propagates towards the resonator by viscous forces.
QCM signals are extremely sensitive to the liposome's height distribution P(y) which depends on 
the actual number and mechanical properties of the tethers, in addition to the usual local attractive/repulsive chemical 
forces. Our approach helps in deciphering the role of hydrodynamics in acoustic sensing and revealing the role of parameters 
hitherto largely unexplored. A practical consequence would be the design of improved biosensors and detection schemes.
\end{abstract}

\maketitle The  quartz crystal microbalance  (QCM) is a  technique for
research at  interfaces with  applications ranging  from nanotribology
and soft matter  to biology, health care  and environmental monitoring
\cite{1a,16,3a,Rodahl97},   covering  a   range   of  length-scales   from
nanometers to  tens of  microns.  In biophysics-related  research, QCM
with dissipation monitoring (QCM-D) operates  in liquids and  follows real-time
changes   in   assemblies   of   lipid   membranes,   DNA,   proteins,
nanoparticles, viruses and cells \cite{Rodahl97,reviakine2011hearing,6a,tsortos2016hydrodynamic,8a,tellechea2009model}.  In a
QCM  experiment, the  analytes form  the interface  between the  solid
substrate (a  quartz crystal  piezoelectric resonator) and  the liquid
environment  (water) where  they are typically exposed  to  $5-150\,
\mathrm{MHz}$ transverse  oscillations; the  sensor monitors,  in real
time,  minute  variations in  the  resonance  frequency $f$  and  energy
dissipation factor $D$ (or decay rate $\Gamma=D f/2$). In vacuum and for
rigid  films the  Sauerbrey  relation \cite{Saue}  indicates that  the
inertia of  the deposited mass  will reduce the  resonator frequency
proportionally to $\Delta f/f$ ; for a resolution of $0.1\ \mathrm{Hz}$
this results  in an  extremely small limit  of detection  of $10^{-12}
\mathrm{g/cm^2}$  .   In  a   liquid,  viscous  forces  propagate  the
resonator's  transverse oscillations  up  to 3  times the  penetration
depth  $\delta=(2\eta/\rho\omega)^{1/2}\in [36,197]\mathrm{nm}$, leading  to
the so-called Stokes flow; here, $\omega=2  \pi f$ while $\eta$ is the
fluid  viscosity and  $\rho$  its density.  In the  case  of a  purely
viscous Newtonian  fluid the shifts $\Delta  f$ and $\Delta D$ were derived
by Kanazawa and Gordon \cite{kanazawa1985frequency}  and subsequent works \cite{12a}.
Johannsmann \cite{13a}  and Voinova \cite{14a}  later used effective-medium   theory   and   phenomenological
constitutive  relations to  estimate  viscoelastic  properties of  the assumed film-formations  of   the  material   covering  the   sensor
surface.
Lately, experiments with discrete particles such as liposomes and viruses made use of a different approach providing estimations of the size of the nanoentities involved \cite{tellechea2009model}. A model developed by Tsortos \cite{tsortos2016hydrodynamic,tsortos2008quantitative,16a} allowed for quantitative size and shape evaluation of biomolecules (DNA, proteins) \cite{17a,mateos2016monitoring};
here, the hydrodynamic quantities of intrinsic viscosity $[\eta]$ and radius
$R_h$ were explicitly taken into account and were linked to the acoustic ratio  $\Delta D/\Delta f$ \cite{tsortos2008quantitative}.  
When the limit of detection for an analyte is of importance (e.g. in biotechnology and in medical applications), the formation of molecular complexes at the surface is one way to enhance the QCM signal.  Signal-enhancers typically of O(100) nm,
such as liposomes, magnetic beads and gold nanoparticles are anchored to the analyte and thus suspended in the fluid, tens of nanometers away from the sensor surface. Viscoelastic film theories based on 1D equations, substantially fail to provide useful information on these and other discrete-particle settings \cite{16,reviakine2011hearing}.

A theoretical understanding of such systems requires a solid hydrodynamic
analysis, involving  3D unstationary flow patterns  resulting from the
viscous  propagation  of  the  fluid-induced  forces  acting  on  the
solutes. Numerical studies of  QCM hydrodynamics of discrete particles
qualitatively reproduce  experimental observations, such  the coverage
dependence of $\Delta D/\Delta f$,  providing preliminary answers to a
large  list of  still  unexplained  phenomena\cite{16}. These  studies
\cite{reviakine2011hearing,tellechea2009model} carried out in 2D using
the commercial  package COMSOL and  more recently in 3D  using lattice
Boltzmann  solvers   \cite{lbqcm,gillissen2018numerical},  considered
adsorbed  rigid particles  at  fixed  positions (obstacles).  However,
fixing the particle  position introduces ad-hoc forces  into the fluid
(those  required  to  keep  the ‘obstacle’  fixed),  which  alter  the
measurement of the system impedance. A correct representation of QCM
hydrodynamics requires solving the dynamics of the analytes, including
the   fluid   traction  acting   on   them   and  other   interactions
(inter-particle forces, surface or contact forces and advection due to
a  mean flow,  if  required); evidently,  this  issue is  particularly
important  in the  case  of  suspended particles. In  this Letter  we
present an experimental  and theoretical study of the  QCM response of
individual liposome-DNA complexes. Simulations  are carried out with a
finite  volume  fluctuating  hydrodynamic  solver  equipped  with  the
immersed  boundary  method, which describes the  particle dynamics and
uses an elastic network  model to reproduce molecular mechanical properties  (e.g. bending
rigidity). The good agreement with  experiments (carried out with POPC
liposomes)  allows  us  to  affirm that  the  analyte  impedance  is
strongly  dominated by  the hydrodynamic  perturbation created  by the
liposome,  which is  suspended in  the liquid  and tethered  by a  DNA
strand (Fig. 1).

{\em Experiments}.
Liposome-DNA (LDNA) complexes were formed by sequential injections of neutravidin (NAv),  DNA,   and  liposomes; more
details in Supplementary   Information  (SI). dsDNA with 21, 50 and 157 base pairs (bp) having lengths $L_{\mathrm{DNA}}$ $7$, $17$ and $53\mathrm{nm}$, respectively, were used.
Liposomes of radius of 15, 25,  50 and 100 nm were considered. In order
to be captured by the NAv  layer previously formed on the surface, DNA
fragments bear a  biotin at one end. In addition,  a cholesterol 
was  incorporated to  the  opposite DNA  end  for subsequent  liposome
binding due to  its strong affinity for the lipid  membrane.
Ring-down QCM  experiments were  performed with an  E4-Qsense (Biolin,
Sweeden)  device  at  $T=25^oC$  under continuous  flow  velocity  of  60
$\mu\mathrm{L/min}$. Measurements  of $\Delta f$ and $\Delta D$ based
on the ring-down  approach have been described elsewhere \cite{QCMD}.
Briefly, after excitation  pulses separated  by milliseconds, a  crystal sensor
resonator  performs underdamped  oscillations described  by
$x(t)  = x_0 \exp[-\Gamma t] \cos[2\pi f t  + \phi]$, where $x_0  \approx 2 \mathrm{nm}$,
and $\Gamma$ and $f$ depend on the
acoustic response related  to the sensor loading. The decay  rate
$\Gamma$ is often expressed in terms of the ``dissipation factor'', $D = 2 \Gamma/f_n$. The fundamental frequency of the particular cut of the quartz crystal is $f_0=5\mathrm{MHz}$
and here we report experimental results for the seveth harmonic $f_7=35 \mathrm{MHz}$.
QCM experiments  monitor  the  time   evolution  of  the  acoustic  signal
registering the  changes in frequency  ($\Delta f$) and dissipation  ($\Delta D$) upon
successive  sample injections  and  surface binding  of  NAv, DNA  and
liposomes (Fig. S1). These shifts increase with  the amount of
deposited  material. To get an intensive quantity 
the procedure consists in plotting the acoustic ratio $-\Delta D/\Delta f$ against $\Delta f$
and extrapolating it to the limit of an infinitesimally small load: this defines
the so called {\em dissipation capacity}  $\ar=-\lim_{\Delta f\rightarrow 0} \Delta D/\Delta f$.
The minus sign is convenient because an extra load usually implies $\Delta f<0$ and $\Delta D>0$ which,
following the analogy with an overdamped spring \cite{16}, are commonly 
intrepreted as extra ``mass'' ($\Delta f$) and ``dissipation'' ($\Delta D$).

{\em Impedance analysis.} Our numerical analysis is based on the well established {\em small load approximation} (SLA) 
\cite{16} which relates the impedance, $\imp = \hat{\sigma}/v_0$ (ratio of the wall stress and the surface velocity) to 
the complex frequency shift $\Delta \tilde{f} = \Delta f + i \Delta \Gamma$ measured in ring-down experiments. The SLA is 
valid if the resonator's mass per unit area is much larger than the load, which is a safe approximation in our case
(where $\Delta f/f_0 \sim 10^{-5}$ or even less).
The impedance of the complex load is expressed as the sum
of  $\imp = \imp^{(Q)} + \impf + \impdna + \impldna$,
where the impedances correspond, respectively, to the clean quartz resonator
$\imp^{(Q)}$,  the (unloaded) Newtonian solvent $\imp^{(0)}$ \cite{kanazawa1985frequency}
the DNA strand {\em without a liposome} $\impdna$ and, the LDNA impedance $\impldna$.
For any contribution (different from $Q$), the SLA yields \cite{16}
$\Delta \tilde{f}^{(a)}/f_0 = \ii \imp^{(a)}/(\pi \imp^{(Q)})$,
where $\imp^{(Q)}=8.8\times 10^{6}\ \mathrm{kg/(m^2 s)}$.
The real part of $\impa$
is related to the dissipation and the imaginary part to the frequency
shift ($\mathrm{Re}[\impa] = \impre \propto -\Delta \Gamma$ and $\mathrm{Im}[\impa] = \impim \propto -\Delta f$) while
the acoustic ratio is $\ar^{(a)}= -2 \impre^{(a)}/(\impim^{(a)} f_n[\mathrm{MHz}])$,
and $f_n$ is the working frequency (here $n=7$).

\begin{figure}
  \begin{center}
    \includegraphics[width=0.3\textwidth]{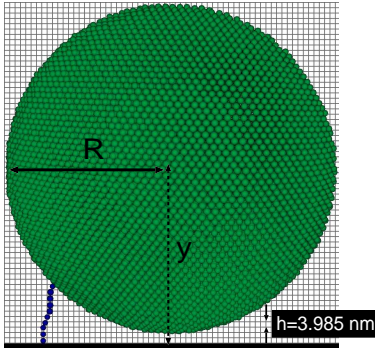}
  \end{center}
  \caption{Snapshot of the numerical model of a liposome of
    $R=100\ \mathrm{nm}$ radius tethered to a DNA strand attached to the resonator wall (see text).
    The coupling between the analyte and the fluid dynamics is
    solved using the immersed boundary method (IBM) \cite{usabiaga2013inertial,fluam}. The spatial
    resolution is set by the fluid cell size $h$ (here $3.958 \mathrm{nm}$).
    The bottom wall oscillates transversally
    at $35 \mathrm{MHz}$,  with a small amplitude (less than one cell).
  \label{fig:1}
  }
\end{figure}

{\em  Simulations.} Our  mesoscopic model  is based  on a  fluctuating
hydrodynamic solver  for compressible unsteady flow equipped  with the immersed
boundary   method    to   couple   fluid   and    structure   dynamics
\cite{usabiaga2013inertial,usabiaga2014inertial,fluam}.      It     is
implemented  in the  GPU  code  {\tt FLUAM}  ({\em  FLuid And  Matter}
interaction),  a  second-order  accurate  finite volume  scheme  on  a
staggered grid  \cite{balboa2012staggered} of  side $h$.   The  liposome  and dsDNA  are  represented
using beads of radius $h$ (Fig.  \ref{fig:1}) connected by harmonic springs and/or bending
potentials  (see SI).  An  elastic network  is
used  to model  the membrane  of the  (hollow) liposome  by connecting
nearest neighbours of  the network with harmonic  springs: the bonding
force  ${\bf  F}_{ij}  =   -k_{L}  ({\bf  r}_{ij}-r_0)$  includes  the
equilibrium distance  $r_0 \approx  h$ and  the spring  constant $k_L$
determining  the liposome  rigidy. Here  we consider  the rigid  limit
(high  $k_L$) to  focus  on the  leading  impedance contribution.  The
double  stranded  DNA  (dsDNA)  is   modelled  by  a  bead  model  for
semiflexible polymers, with bending energy extracted
from the DNA persistence length at  room temperature (50 nm).
We  shall  use the  term  {\em  link}  to  denote the  DNA-wall  force
$F_{\mathrm{link}}$.

The geometry is illustrated in Fig. \ref{fig:1}.
The simulation box is periodic in the resonator plane $x-z$, with
dimensions $L\times L_y \times L$. Rigid no-slip walls
at $y=0$ and $y=L_y$ are imposed via explicit boundary conditions \cite{usabiaga2013inertial,florentesis}). The top wall is kept at rest while the bottom wall at $y=0$ moves in the $x$ direction with velocity $v_{wall}=v_0 \cos(2\pi f t)$
with $v_0$ set to fix a small wall displacement $x_0 < h$.
To achieve the required numerical convergence we used
a spatial resolution of $h=3.958\ \nm$ (see SI).
The code units map the density and kinematic viscosity of water at $T=25^o\,\mathrm{C}$ (see SI). The sound velocity $c$ was set to match the experimental value
of the group $f_7\delta/c \sim 3\times 10^{-3}$, whose smallness
indicates a minor effect of fluid-compressibility \cite{mazur1974generalization}. Moreover, the large time scale separation between liposome diffusion time and the QCM oscillation period ($6\pi \eta R^3 f/k_BT > 10^4$) makes it possible to neglect thermal fluctuations over
the short simulation runs we used to evaluate the impedance (about 10 periods).
We stress, however, that in our simulations the liposome is free to move according to
the flow traction; this is essential to obtain unbiased results,
particularly because the liposomes are {\em not} adsorbed but suspended.

  {\em Computational protocol.}  An ``instantaneous'' value of $f$ and
  $D$ in  the QCM experiment  corresponds to an ensemble  average over
  many   LDNA  assemblies  over a large number of  pulse-waiting time sequences
  (each one in the  milisecond range).  We  average over  a set of  40 initial
  equilibrium  configurations extracted  from Monte Carlo sampling
  of the LDNA complex (SI  for details).  The
  impedance of each configuration is measured running {\tt FLUAM} over
  10  periods, after the transient regime. We  record the  time-depedent wall stress  $\sigma(t) =
  \sigma_{\mathrm{hydro}} + \sigma_{\mathrm{link}}$ with contributions
  from                the                 DNA-wall                link
  $\sigma_{\mathrm{link}}=F_{\mathrm{link}}/L^2$    and    the    wall
  hydrodynamic  stress  $\sigma_{\mathrm{hydro}}   (t)=  \eta  \langle
  (\partial v_x({\bf r},t)/\partial  y)_{wall}\rangle$ (angle brackets denote
  average   over   the   whole   surface).    Setting   $\sigma(t)   =
  \mathrm{Re}[\hat{\sigma} \exp(-\ii\omega  t)]$ we obtain  the stress
  phasor    $\hat{\sigma}$     by    fitting    to     $\sigma(t)    =
  \mathrm{Re}[\hat{\sigma}] \cos(\omega t) + \mathrm{Im}[\hat{\sigma}]
  \sin(\omega   t)$.   This   provides  the   {\em  total}   impedance
  $\imp=\hat{\sigma}/v_0$   of   each  initial   configuration.    The
  LDNA impedance  is $\impldna= \imp-\impf -\impdna$, where
  $\impf =(\ii-1)\,\eta/\delta$ corresponds to the  unperturbed   (Stokes)
  flow
  and $\impdna$ to the DNA anchor (without liposome)
  which was found to be negligible.

Calculations of $\impldna$ were
carried out for increasing box side $L$ (the coverage being $\phi=1/L^2$).
Although a periodic array of a single analyte is far
from the experimental randomness, the variation of $\Delta D/\Delta f$ with $\phi$
is similar to that observed in experiments (see SI).
Consistently, in simulations, the dissipative capacity is defined as the value of
the acoustic ratio in the limit of low coverage 
$\ar_{num} = -\lim_{\phi \rightarrow 0} 2 \mathrm{Re}[\impldna]/(\mathrm{Im}[\impldna] f_n)$.

\begin{figure}
    \begin{center}     
      \includegraphics[width=0.45\textwidth]{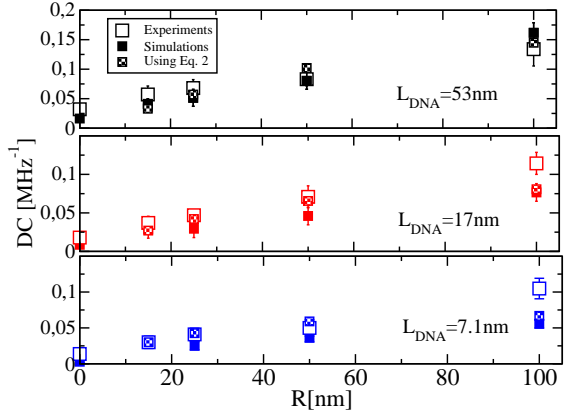}\\
    \end{center}
    \caption{Dissipative capacity
of liposome-DNA complexes at frequency $f_n=35\ \mathrm{MHz}$,
against the liposome radius $R$
for different dsDNA contour lengths
Filled symbols correspond to
simulation results, crossed symbols to Eq. \ref{eq:avez}
(using Eq. \ref{eq:z} for $\impfree$)
and open symbols to experiments using POPC liposomes at $T=25^o$C
\cite{electra2020}. Simulations were performed with a mesh of $h=3.958\ \mathrm{nm}$ in boxes  of $L=506.6\ \mathrm{nm}$ side.}
  \label{fig:ar}
\end{figure}

{\em Results.}
Numerical and experimental estimations of the acoustic ratio
$\ar\equiv \ar^{\mathrm{(LDNA)}}$ are compared in Fig. \ref{fig:ar},
both showing an increase of $\ar$  with the liposome radius $R$ and the DNA contour length $L_{\mathrm{DNA}}$. The agreement is quite close to being quantitative and gives us confidence to deploy our model in future efforts
to gradually understand the large number of factors 
affecting the value of $\ar$ (some of them will be mentioned below).
Such a task demands a theoretical  analysis of the impedance, presented below.

\begin{figure}
  \begin{center}
    \includegraphics[width=0.4\textwidth]{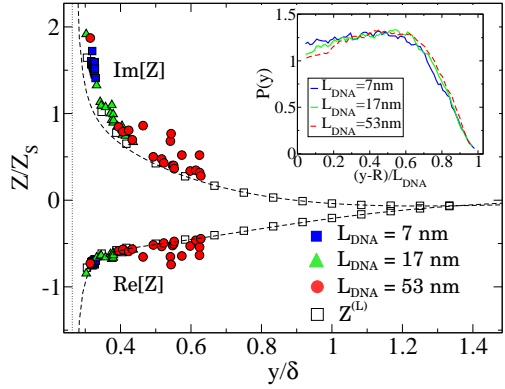}\\    
  \end{center}
  \caption{Impedance of a liposome of $R=25\ \mathrm{nm}$ versus the scaled wall distance
    $y/\delta$ (penetration length  $\delta=95\,\mathrm{nm}$ for  $f_7=35\ \mathrm{MHz}$). Open squares correspond to a freely suspended liposome ($\impfree$) and filled circles to individual configurations of the liposome-DNA complex for several contour lengths. Results obtained in a square box of side  $L=253,3\ \mathrm{nm}$ and mesh-size (resolution) $h=3.958\ \mathrm{nm}$. Dashed lines correspond to Eq. \ref{eq:z} (the vertical line indicates the wall). Inset: Monte Carlo results for the probability density of finding the liposome ($R=25\ \mathrm{nm}$) at height $y$. The scaled length $(y-R)/L_{\mathrm{DNA}}$ provides a master curve for all values of $L_{\mathrm{DNA}}$ (see SI).
  }
  \label{fig:z}
\end{figure}

{\em Analysis.}  The impedance of the LDNA assembly can be decomposed
into  contributions from  the liposome  $(L)$  and the  DNA $\impldna  =
\impldna_L+\impldna_{\mathrm{DNA}}$.   The term  $\impldna_{\mathrm{DNA}}$ collects  the
contributions of the linker and the hydrodynamic impedance of the DNA.
Note that  with the  exception of the  small molecular  linker (biotin),
the  DNA chain is suspended in  the fluid and the dominant part of its impedance is
of hydrodynamic origin. In general, the
hydrodynamic impedance arises from  the fluid-induced forces (or force
distributions) on the solutes, which  are transfered {\em back} to the
fluid by momentum  conservation and propagate by  viscous diffusion to
the resonator, creating stress on the wall. The lack of symmetry makes elusive an
analytical approach to the fluid-induced dynamics, even for a point-particle under
a Stokes flow \cite{morrison2018}. Fluid-induced forces could be of
inertial origin  (relative fluid-particle  accelerations leading  to a
net force  on the solute). However,  due to the extremely small density constrast of
the \textit{quasi-neutrally-buoyant} liposomes and DNA, inertia is
negligible. Even  so, liposomes will  bear a
stress distribution  at their  surface due  to their  reaction against
flow-induced deformation.  The induced liposome surface  stress, along
with the line  tension along the DNA, are propagated  back to the wall
by  viscous  transport. There,  at  the  resonator,  the only  way  to
disentangle  $\impldna_L$  from  $\impldna_{\mathrm{DNA}}$ is  to  compare  the
impedance of  a free-floating (not tethered)  liposome $\impfree$ with
that of  a LDNA assembly  where the liposome  is placed at  a similar
height $y$.  This comparison, in  Fig. \ref{fig:z}, clearly shows that
the  dominant contribution  to  $\impldna$ comes  from the  liposome's
surface stress, determined by $\impfree(y)$. Before analyzing the role
of the     DNA    strand,     we     focus     on    $\impfree(y)$     in
Fig. \ref{fig:z}. Remarkably $\impfree(y)$ decays almost exponentially
with $y$: this general trend can be explained by hydrodynamic reflection.
Resonator vibrations  (with velocity $v_0$) are  propagated upwards to
the   fluid  by   viscous  forces   and  creates   the  unsteady   and
distance-dependent  Stokes flow.   Its velocity  field (phasor),  $v_0
\exp[-\alpha y]$, contains the viscous propagator ($\exp[-\alpha y]$, with $\alpha = (1-\ii)/\delta$).
The behavior of $\impfree(y)$ can be understood using a
corollary of Fax\'en's theorem (valid for steady \cite{Kim}
and unsteady flow \cite{poz}), which guarantees that
the solute's surface stress is a linear function of the ambient flow.
Far away from the resonator, the ambient flow is similar to the Stokes flow,
so the liposome surface stress should scale as $\exp[-y\alpha]$. The stress
reflects {\em back} to the surface by viscous forces leading to
$\impfree \sim Z_S \exp[-2 \alpha y]$, where the prefactor $Z_S=(20\pi/3) \eta R^3/(L^2\delta^2)$
is taken to be consistent with the stresslet of a sphere under steady shear
\footnote{The steady stresslet is $\mathcal{S} = \frac{20}{3} \pi \eta R^3 \left(1 + \frac{R^2}{10} \nabla^2\right) \mathcal{E}^{(0)}$
  and take the strain component $\mathcal{E}^{(0)}_{xy}=(1/2) dv^{(0)}_x/dy$.}.  Close to the resonator,
the ambient flow includes a significant contribution
from the wall reaction field \cite{felderhof2012hydrodynamic} and the impedance
becomes a decreasing function of the surface-to-sphere distance $y-R$. This reasoning
leads us to the following ansatz,
\begin{equation}
  \label{eq:z}
  \impfree(y) = Z_S \left[(A+\ii B) \exp[-2 \alpha y] + \frac{2 \ii C }{\alpha(y-R)}\right].
\end{equation}
Using $A \approx 1.40 (R/\delta)^2$, $B\approx 1.5-0.03\exp(2.5\,R/\delta)$ and $C\approx 0.01$ (see SI), Eq. \ref{eq:z} fits our numerical results for $y/\delta >0.04$  \footnote{We have verified that the divergence saturates at contact.}
within less than $5\%$ error (see Fig. \ref{fig:z}).

The contribution of the DNA can now be estimated as $\impldna_{\mathrm{DNA}}= \impldna - \impfree$.
As shown in Fig. S5,  $\impldna_{\mathrm{DNA}}$ mildly increases with $L_{\mathrm{DNA}}$. Its imaginary part does not greatly vary with $R$ while its real part increases linearly with $R$. We find that the DNA represents a minor  contribution to the total impedance $\impldna$ for $R>50\mathrm{nm}$, while it becomes noticeable for smaller liposomes (see SI).

Figure \ref{fig:z} shows that that the dispersion of $\impldna$
around its average decaying trend is relatively small and
does not greatly vary with $y$. Thus $\impldna_{\mathrm{DNA}}$
does not strongly vary with the orientation of the DNA strand. The main effect of the DNA
is to constrain the liposome position and, notably, to determine its height distribution $P(y)$.
The average impedance is obtained from the weighted average,
\begin{equation}
  \label{eq:avez}
  \langle \impldna \rangle = \int P(y) \impldna(y) dy,
\end{equation}
where $\impldna(y)= \impfree(y) + \impldna_{\mathrm{DNA}}$.
Relation \ref{eq:avez} is extremely useful because it decomposes the analyte and anchor contributions, allowing for a fast evaluation of non-trivial effects.
$P(y)$ encodes important microscopic information about
the anchor: bending rigidity, linker-DNA tilt energy \cite{pettit2004} or large values of the DNA coverage \cite{ling2010}
which could lead to multiple anchors connected to the liposome \cite{electra2020}).
In general, $P(y)$ can also introduce information into Eq. \ref{eq:avez}
about physico-chemical forces between the analyte and the wall (solvation, dispersion, electrostatic forces) as well as the effect of advection under a strong Poiseulle flow.
All these effects are known to alter the acoustic ratio and their relevance can be tested using Eq. \ref{eq:avez}, by pre-evaluating $P(y)$, either theoretically or from Monte Carlo simulations (MC). We use MC sampling to obtain $P(y)$ 
for an electrically neutral liposome (POPC) anchored by a single DNA chain. The inset of Fig. \ref{fig:z} shows $P(y)$, which applied to Eq. \ref{eq:avez}
(fed with $\impfree$ in Eq. \ref{eq:z} and $\impdna_{\mathrm{DNA}}$) predicts values of DC quite close to experimental results (crossed symbols in Fig. \ref{fig:ar}).

We have shown that the acoustic response of suspended particles in QCM is mainly determined by the hydrodynamic response of the analyte,
which strongly depends on its height distribution $P(y)$. The hydrodynamic response of liposomes also depends
on their bending rigidity $\kappa$ and membrane fluidity \cite{Kim}.
Soft liposomes, with smaller $\kappa$ and higher fluidity,
experimentally yield slightly smaller DC \cite{electra2020}. A recent analytical study \cite{marc} indicates that both effects
might be antagonistic (DC decreases with $\kappa$  but increases with fluidity). Disentangling all this subtle information hidden beneath the complex QCM hydrodynamic fields requires a combined effort of experiments, simulations and hydrodynamic theory. This work represents a step in this direction.

{\em Acknowledgments.} We acknowledge the EU Commision for the FET-OPEN CATCH-U-DNA project.
R.D-B acknowledges support from MINECO via the proyect FIS2017-86007-C3-1-p.


\end{document}